\documentclass[preprint,aip,jmp,amsmath,amssymb,reprint,]{revtex4}     % one column

\usepackage{textcase}
\usepackage{graphicx}% Include figure files
\usepackage{dcolumn}% Align table columns on decimal point
\usepackage{bm}% bold math
\usepackage{amsmath,amsthm,amssymb,mathrsfs}

\begin{document}

\title{Magnetoresistance originated from charge-spin conversion in ferromagnet}

\author{Tomohiro Taniguchi}
\email{tomohiro-taniguchi@aist.go.jp}

\affiliation{ 
National Institute of Advanced Industrial Science and Technology (AIST), Spintronics Research Center, Tsukuba 305-8568, Japan}

\date{\today}% 

\begin{abstract}
Transverse magnetoresistance in a ferromagnetic/nonmagnetic/ferromagnetic trilayer originated from charge-spin conversion by anomalous Hall effect is investigated theoretically. 
Solving the spin diffusion equation in bulk and using the spin-dependent Landauer formula at the ferromagnetic/nonmagnetic interface, 
an analytical formula of the transverse resistivity is obtained. 
The charge-spin conversion by the anomalous Hall effect contributes to the magnetoresistance in a manner proportional to the square of the spin anomalous Hall angle. 
The angular dependence of the magnetoresistance is basically identical to that of planar Hall effect, 
but has an additional term which depends on the relative angle of the magnetizations in two ferromagnets. 
\end{abstract}

\maketitle

% ===================================================================================================================================================================================== %

\section{Introduction}
\label{sec:Introduction}

Spin Hall effect originated from spin-orbit interaction in nonmagnetic heavy metal has attracted much attention from both fundamental and applied physics 
\cite{dyakonov71,hirsch99,kato04}. 
The spin Hall effect generates electric (charge) currents from spin currents and \textit{vice\ versa} \cite{dyakonov07,takahashi08,hoffmann13,saslow15,taniguchi16APEX}, 
which is called charge-spin conversion. 
The charge-spin conversion in a nonmagnet has been studied experimentally in ferromagnetic/nonmagnetic multilayers by several methods, 
such as nonlocal spin current diffusion \cite{kimura07,seki08}, 
the combination of the inverse spin Hall effect and spin pumping \cite{ando08,mosendz10}, 
spin-torque ferromagnetic resonance \cite{liu11,kondou12}, 
first and second harmonic Hall voltage \cite{garello13,kim13}, 
and spin Hall magnetoresistance \cite{nakayama13,althammer13,kim16}. 

% ===================================================================================================================================================================================== %

The spin-orbit interaction in a ferromagnet gives anomalous Hall effect \cite{pugh53,nagaosa10}, 
where an electric voltage is generated in the direction perpendicular to both the magnetization and an external electric field. 
Note that the current generated by the anomalous Hall effect is spin-polarized. %because of the spin-dependent transport properties in ferromagnets. 
Recently, a theory of spin current generation and spin-torque excitation by the anomalous Hall effect was proposed \cite{taniguchi15}. 
Magnetoresistance effects originated from the charge-spin conversion by the anomalous Hall effect were also predicted 
in a single ferromagnet \cite{taniguchi16SPIE} and ferromagnetic/nonmagnetic bilayer \cite{taniguchi16PRB}. 
Very recently, on the other hand, the observations of the charge-spin conversion by the anomalous Hall effect, or related phenomena, 
have been reported in a multilayer including two ferromagnets \cite{miao13,wang14,tsukahara14,wahler16,comment1,comment2}. 

% ===================================================================================================================================================================================== %

In this paper, we investigate the magnetoresistance effect in a ferromagnetic/nonmagnetic/ferromagnetic trilayer originated from the charge-spin conversion by the anomalous Hall effect. 
By solving the diffusion equation of spin accumulation and the spin-dependent Landauer formula, 
an analytical formula of the transverse resistivity is derived. 
%It is shown that the charge-spin conversion by the anomalous Hall effect gives magnetoresistances on the order of the square of the spin anomalous Hall angle. 
In addition to the magnetoresistance found in previous works, where the angular dependence is the same with the planar Hall effect, 
a contribution to the magnetoresistance which depends on the relative angle of the magnetizations in two ferromagnets is revealed. 

% ===================================================================================================================================================================================== %

The paper is organized as follow. 
In Sec. \ref{sec:System description}, we describe the system studied in this paper. 
In Sec. \ref{sec:Magnetoresistance in ferromagnetic/nonmagnetic trilayer}, 
the transverse magnetoresistance in the ferromagnetic/nonmagnetic/ferromagnetic trilayer is calculated. 
The conclusion is summarized in Sec. \ref{sec:Conclusion}.

% ===================================================================================================================================================================================== %

% ===================================================================================================================================================================================== %

\section{System description}
\label{sec:System description}

The system we consider is a ferromagnetic(F${}_{1}$)/nonmagnetic(N)/ferromagnetic(F${}_{2}$) trilayer structure shown in Fig. \ref{fig:fig1}(a). 
We denote the unit vector pointing in the magnetization direction in the F${}_{k}$ ($k=1,2$) layer as $\mathbf{m}_{k}$ and the thickness of the layer as $d_{k}$. 
An external electric field is applied along $x$ direction.
We assume that the F${}_{2}$ layer shows the anomalous Hall effect, 
and injects spin currents into the F${}_{1}$ layer placed along $z$ direction. 
The anomalous Hall effect also provides electric currents flowing in $y$ direction given by \cite{taniguchi16PRB} 
\begin{equation}
\begin{split}
  J_{{\rm c}y}
  &=
  \sigma_{\rm AH}
  \left[
    m_{2z}
    -
    \left(
      \frac{\sigma_{\rm AH}}{\sigma_{\rm F}}
    \right)
    m_{2x}
    m_{2y}
  \right]
  E_{x}
\\
  &+
  \frac{(\beta-\zeta) \sigma_{\rm AH}}{e}
  m_{2x}
  \partial_{z}
  \delta
  \mu_{\rm F_{2}}. 
  \label{eq:Jcy_def}
\end{split}
\end{equation}
Here, $\sigma_{\rm F}$ and $\sigma_{\rm AH}$ are the longitudinal and transverse (anomalous Hall) conductivities, respectively. 
The spin polarization of these conductivities in respective are $\beta$ and $\zeta$ \cite{taniguchi15}. 
The spin accumulation defined from an electrochemical potential $\bar{\mu}_{s}$ of the spin-$s$ ($s=\uparrow,\downarrow$) electrons is 
denoted as $\delta\mu=(\bar{\mu}_{\uparrow}-\bar{\mu}_{\downarrow})/2$. 

% ===================================================================================================================================================================================== %

The first and second terms on the right hand side of Eq. (\ref{eq:Jcy_def}) are the electric currents generated from the anomalous Hall effect. 
On the other hand, the last term represents a contribution from the charge-spin conversion by the anomalous Hall effect \cite{taniguchi16SPIE,taniguchi16PRB}. 
A factor 
\begin{equation}
  \vartheta 
  \equiv 
  \frac{(\beta-\zeta) \sigma_{\rm AH}}{\sigma_{\rm F}}, 
  \label{eq:spin_Hall_angle}
\end{equation}
characterizes the efficiency of the charge-spin conversion by the anomalous Hall effect. 
Therefore, let us define Eq. (\ref{eq:spin_Hall_angle}) as the spin anomalous Hall angle. 
The physical meaning of the spin anomalous Hall angle is as follows. 
A factor $\sigma_{\rm AH}/\sigma_{\rm F}$ is the anomalous Hall angle \cite{nagaosa10}, 
characterizing the ratio of the transverse voltage to the external voltage. 
%In other words, the anomalous Hall angle is determined by the amount of the electrons scattered to the transverse direction by the spin-orbit interaction. 
The spin polarization of the transverse current is given by the product of the anomalous Hall angle and the spin polarization, $\zeta$, 
i.e., $\zeta (\sigma_{\rm AH}/\sigma_{\rm F})$. 
Note that the electrons scattered to the transverse direction creates the charge accumulation. 
%Here, we use the open-circuit boundary condition to the transverse direction. 
The electrons then move along the direction of an internal electric field generated by the charge accumulation. 
The electric current due to this internal field is also spin polarized, where the spin polarization is given by $\beta$. 
The net spin polarization decreases because of this motion of the electrons. 
As a result, the total spin polarization of the spin current generated by the anomalous Hall effect is given by 
$|(\beta-\zeta)\sigma_{\rm AH}/\sigma_{\rm F}|$, which is the spin anomalous Hall angle defined in Eq. (\ref{eq:spin_Hall_angle}). 

% ===================================================================================================================================================================================== %

In experiments related to spin Hall effect such as, for example the measurements of the harmonic Hall voltage \cite{kim13,comment2}, 
the spin current was measured from the planar Hall effect in the transverse direction. 
The angular dependence of the planar Hall effect is described by $m_{x}m_{y}$. 
As can be seen in Eq. (\ref{eq:Jcy_def}) and the results shown in our previous work \cite{taniguchi16PRB}, 
the higher order term of the anomalous Hall effect and the contribution from the charge-spin conversion, 
corresponding to the second and third terms of Eq. (\ref{eq:Jcy_def}), respectively, 
have the same angular dependence as $m_{x}m_{y}$. 
We should emphasize here that our previous work \cite{taniguchi16PRB} focuses on a ferromagnetic/nonmagnetic bilayer. 
On the other hand, as will be shown in the next section, 
the present trilayer has another contribution to the magnetoresistance whose angular dependence is described by 
not only by $m_{x}m_{y}$ but also by $\mathbf{m}_{1}\cdot\mathbf{m}_{2}$. 

% ===================================================================================================================================================================================== %

\section{Magnetoresistance in ferromagnetic/nonmagnetic trilayer}
\label{sec:Magnetoresistance in ferromagnetic/nonmagnetic trilayer}

The spin accumulation $\delta\mu_{\rm F}$ in the F${}_{2}$ layer should be evaluated to calculate the magnetoresistance from Eq. (\ref{eq:Jcy_def}). 
The spin accumulations obey the diffusion equation \cite{takahashi08,valet93}, 
\begin{equation}
  \frac{\partial^{2} \delta\mu_{\rm F}}{\partial z^{2}}
  =
  \frac{\delta \mu_{\rm F}}{\ell^{2}}, 
  \label{eq:diffusion_equation}
\end{equation}
where $\ell$ is the spin diffusion length. 
The spin accumulation in the F${}_{2}$ layer is related to the spin current density in the F${}_{2}$ layer via \cite{taniguchi15} 
\begin{equation}
\begin{split}
  J_{{\rm s}i}
  =
  &
  -\frac{\hbar\sigma_{\rm F}}{2e^{2}}
  \partial_{i}
  \delta
  \mu_{\rm F_{2}}
  -
  \frac{\hbar \sigma_{\rm AH}}{2e^{2}}
  \epsilon_{i\alpha k}
  m_{\alpha}
  \partial_{k}
  \delta
  \mu_{\rm F_{2}}
\\
  &-
  \frac{\hbar \beta \sigma_{\rm F}}{2e^{2}}
  \partial_{i}
  \bar{\mu}_{\rm F_{2}}
  -
  \frac{\hbar \zeta \sigma_{\rm AH}}{2e^{2}}
  \epsilon_{i \alpha k}
  m_{\alpha}
  \partial_{k}
  \bar{\mu}_{\rm F_{2}}, 
\end{split}
\end{equation}
where $\bar{\mu}=(\bar{\mu}_{\uparrow}+\bar{\mu}_{\downarrow})/2$, 
and $\epsilon_{ijk}$ is the Levi-Civita asymmetric tensor. 
The spin current density in the F${}_{1}$ layer is obtained in a similar way by neglecting terms related to $\sigma_{\rm AH}$. 
The boundary conditions of Eq. (\ref{eq:diffusion_equation}) are given by the spin currents at the boundaries. 
For the F${}_{2}$ layer, the spin current is zero at the outer boundary, $z=0$. 
On the other hand, we denote the spin current at the F${}_{2}$/N interface as $\mathbf{J}_{\rm s}^{\rm F_{2} \to N}$. 
For simplicity, we assume that the penetration depth of the transverse spin current in the F${}_{2}$ layer is sufficiently short \cite{slonczewski96,stiles02,zhang02,zhang04,taniguchi08,taniguchi08PRB}, 
and therefore, only the component of $\mathbf{J}_{\rm s}^{\rm F_{2} \to N}$ parallel to $\mathbf{m}_{2}$ survives inside the F${}_{2}$ layer. 
Consequently, the solution of $\delta\mu_{\rm F}$ in the F${}_{2}$ layer is given by 
\begin{equation}
\begin{split}
  \delta
  \mu_{\rm F_{2}}
  =
  \frac{4\pi e^{2} \ell}{(1-\beta^{2}) h \sigma_{\rm F} \sinh(d_{2}/\ell)}
  &
  \left\{
    \frac{\hbar(\beta-\zeta)\sigma_{\rm AH}}{2e}
    m_{2y}
    E_{x}
    \left[
      \cosh
      \left(
        \frac{z-d_{2}}{\ell}
      \right)
      -
      \cosh
      \left(
        \frac{z}{\ell}
      \right)
    \right]
  \right.
\\
  &
  \left.
    -
    \mathbf{m}_{2}
    \cdot
    \mathbf{J}_{\rm s}^{\rm F_{2} \to N}
    \cosh
    \left(
      \frac{z}{\ell}
    \right)
  \right\}.
  \label{eq:spin_accumulation_F2}
\end{split}
\end{equation}
The spin accumulation in the F${}_{1}$ layer is obtained in a similar way. 
The spin current at the F/N interface is given by the spin-dependent Landauer formula \cite{brataas01}, 
\begin{equation}
  \mathbf{J}_{\rm s}^{{\rm F} \to {\rm N}}
  =
  \frac{1}{2\pi S} 
  \left[
    \frac{(1-\gamma^{2}) g}{2}
    \mathbf{m}
    \cdot
    \left(
      \delta
      \bm{\mu}_{\rm F}
      -
      \delta
      \bm{\mu}_{\rm N}
    \right)
    \mathbf{m}
    -
    g_{\rm r}
    \mathbf{m}
    \times
    \left(
      \delta
      \bm{\mu}_{\rm N}
      \times
      \mathbf{m}
    \right)
  \right], 
  \label{eq:spin_current_FN}
\end{equation}
where $g$ and $\gamma$ are the dimensionless interface conductance and its spin polarization, respectively. 
The conductance $g$ is related to the interface resistance $r$ via $r=(h/e^{2})S/g$, where $S$ is the cross-section area. 
The real part of the mixing conductance is denoted as $g_{\rm r}$, 
whereas that of the imaginary part is assumed to be zero, for simplicity \cite{zwierzycki05}. 
The spin accumulation in the ferromagnet is $\delta\bm{\mu}_{\rm F}=\delta\mu_{\rm F}\mathbf{m}$, 
whereas $\delta\bm{\mu}_{\rm N}$ is the spin accumulation in the nonmagnet. 
Substituting Eq. (\ref{eq:spin_accumulation_F2}) into Eq. (\ref{eq:spin_current_FN}), 
the spin current at the F${}_{2}$/N interface is rewritten as 
\begin{equation}
\begin{split}
  \mathbf{J}_{\rm s}^{\rm F_{2} \to N}
  =&
  -\frac{\hbar g^{*} (\beta-\zeta) \sigma_{\rm AH}}{2eg_{\rm F}}
  m_{2y}
  \tanh
  \left(
    \frac{d_{2}}{2\ell}
  \right)
  E_{x} 
  \mathbf{m}_{2}
\\
  &-
  \frac{1}{2\pi S}
  \left[
    g^{*}
    \left(
      \mathbf{m}_{2}
      \cdot
      \delta 
      \bm{\mu}_{\rm N}
    \right)
    \mathbf{m}_{2}
    +
    g_{\rm r}
    \mathbf{m}_{2}
    \times
    \left(
      \delta
      \bm{\mu}_{\rm N}
      \times
      \mathbf{m}_{2}
    \right)
  \right],
  \label{eq:spin_current_F2N}
\end{split}
\end{equation}
where $g_{\rm F}$ and $g^{*}$ are defined as 
\begin{equation}
  \frac{g_{\rm F}}{S}
  =
  \frac{h(1-\beta^{2})\sigma_{\rm F}}{2e^{2}\ell}, 
\end{equation}
\begin{equation}
  \frac{1}{g^{*}}
  =
  \frac{2}{(1-\gamma^{2})g}
  +
  \frac{1}{g_{\rm F} \tanh(d_{2}/\ell)}. 
\end{equation}
The spin current at the F${}_{1}$/N interface is obtained in a similar way. 
%by putting $\sigma_{\rm AH}$ in Eq. (\ref{eq:spin_current_F2N}) zero. 

% ===================================================================================================================================================================================== %

We assume that the thickness of the nonmagnet is sufficiently thinner than its spin diffusion length, as usually adopted in experiments \cite{comment1}. 
Thus, the spin current in the nonmagnet is conserved, i.e., 
$\mathbf{J}_{\rm s}^{\rm F_{1} \to N}+\mathbf{J}_{\rm s}^{\rm F_{2} \to N}=\bm{0}$. 
Then, the spin accumulation in the nonmagnet is given by \cite{taniguchi15} 
\begin{equation}
\begin{split}
  \delta 
  \bm{\mu}_{\rm N}
  =&
  -\frac{2\pi \hbar [1+\lambda (\mathbf{m}_{1}\cdot\mathbf{m}_{2})^{2}] g^{*} (\beta-\zeta) \sigma_{\rm AH}}{2eg_{\rm F}(g_{\rm r}+g^{*})[1-\lambda^{2}(\mathbf{m}_{1}\cdot\mathbf{m}_{2})^{2}]}
  \tanh
  \left(
    \frac{d_{2}}{2 \ell_{\rm F}}
  \right)
  m_{2y}
  E_{x} S
  \mathbf{m}_{2}
\\
  &-
  \frac{2\pi \hbar \lambda (\mathbf{m}_{1}\cdot\mathbf{m}_{2}) g^{*} (\beta-\zeta) \sigma_{\rm AH}}{2eg_{\rm F}(g_{\rm r}+g^{*})[1-\lambda^{2}(\mathbf{m}_{1}\cdot\mathbf{m}_{2})^{2}]}
  \tanh
  \left(
    \frac{d_{2}}{2 \ell_{\rm F}}
  \right)
  m_{2y}
  E_{x} S
  \mathbf{m}_{2}
  \times
  \left(
    \mathbf{m}_{1}
    \times
    \mathbf{m}_{2}
  \right).
  \label{eq:spin_accumulation}
\end{split}
\end{equation}
Here, we define $\lambda=(g_{\rm r}-g^{*})/(g_{\rm r}+g^{*})$. 
We note from Eq. (\ref{eq:spin_current_F2N}) that 
\begin{equation}
  \mathbf{m}_{2}
  \cdot
  \mathbf{J}_{\rm s}^{\rm F_{2} \to N}
  =
  -\frac{\hbar g^{*}(\beta-\zeta) \sigma_{\rm AH}}{2eg_{\rm F}}
  \tanh
  \left(
    \frac{d}{2\ell}
  \right)
  E_{x}
  -
  \frac{g^{*}}{2\pi}
  \mathbf{m}_{2}
  \cdot
  \delta
  \bm{\mu}_{\rm N}.
  \label{eq:spin_current_F2N_long}
\end{equation}
Substituting Eq. (\ref{eq:spin_current_F2N_long}) together with Eq. (\ref{eq:spin_accumulation}) into Eq. (\ref{eq:spin_accumulation_F2}), 
the solution of $\delta\mu_{\rm F_{2}}$ is obtained. 

% ===================================================================================================================================================================================== %

From Eq. (\ref{eq:Jcy_def}), we define the averaged current density in the $y$ direction as 
\begin{equation}
\begin{split}
  \overline{J_{{\rm c}y}}
  &\equiv
  \frac{1}{d_{2}}
  \int_{0}^{d_{2}}
  d z 
  J_{{\rm c}y}
\\
  &=
  \sigma_{\rm AH}
  \left[
    m_{2z}
    -
    \left(
      \frac{\sigma_{\rm AH}}{\sigma_{\rm F}}
    \right)
    m_{2x}
    m_{2y}
  \right]
  +
  \frac{(\beta-\zeta) \sigma_{\rm AH}}{ed_{2}}
  m_{2x}
  \left[
    \delta
    \mu_{\rm F_{2}}(z=d_{2})
    -
    \delta
    \mu_{\rm F_{2}}(z=0)
  \right].
\end{split}
\end{equation}
We also define the transverse resistivity as $\rho^{\rm T}=-(\overline{J_{{\rm c}y}}/E_{x})/\sigma_{\rm F}^{2}$ \cite{chen13}. 
Using Eqs. (\ref{eq:spin_accumulation_F2}), (\ref{eq:spin_accumulation}), and (\ref{eq:spin_current_F2N_long}), 
we find that the transverse resistivity is given by 
\begin{equation}
  \rho^{\rm T}
  =
  -\rho_{\rm F}
  \left(
    \frac{\sigma_{\rm AH}}{\sigma_{\rm F}}
  \right)
  m_{2z}
  +
  \rho_{\rm F}
  \left(
    \frac{\sigma_{\rm AH}}{\sigma_{\rm F}}
  \right)^{2}
  m_{2x}
  m_{2y}
  +
  \left[
    \Delta
    \rho_{1}
    +
    \Delta
    \rho_{2}(\mathbf{m}_{1},\mathbf{m}_{2})
  \right]
  m_{2x}
  m_{2y},
  \label{eq:total_resistivity}
\end{equation}
where $\rho_{\rm F}=1/\sigma_{\rm F}$ is the resistivity. 
The first two terms in Eq. (\ref{eq:total_resistivity}) are the conventional transverse resistivities resulting from the anomalous Hall effect. 
On the other hand, the last two terms originate from the charge-spin conversion by the anomalous Hall effect. 
The term $\Delta\rho_{1}$ is the resistivity found in our previous work \cite{taniguchi16SPIE,taniguchi16PRB} given by
\begin{equation}
\begin{split}
  \frac{\Delta\rho_{1}}{\rho_{\rm F}}
  =&
  \frac{\ell}{(1-\beta^{2})d_{2}}
  \left[
    \frac{(\beta-\zeta)\sigma_{\rm AH}}{\sigma_{\rm F}}
  \right]^{2}
  \left[
    2
    -
    \frac{g^{*}}{g_{\rm F}}
    \tanh
    \left(
      \frac{d_{2}}{2\ell}
    \right)
  \right]
  \tanh
  \left(
    \frac{d_{2}}{2\ell}
  \right).
\end{split}
\end{equation}
On the other hand, $\Delta\rho_{2}$ in Eq. (\ref{eq:total_resistivity}) is a new term found in this study. 
This resistivity depends on the relative angle of the magnetizations in the F${}_{1}$ and F${}_{2}$ layers through the term $\mathbf{m}_{1}\cdot\mathbf{m}_{2}$, 
and is given by 
\begin{equation}
\begin{split}
  \frac{\Delta\rho_{2}}{\rho_{\rm F}}
  =
  -\frac{\ell}{(1-\beta^{2})d_{2}}
  \left[
    \frac{(\beta-\zeta)\sigma_{\rm AH}}{\sigma_{\rm F}}
  \right]^{2}
  \frac{g^{*}}{g_{\rm F}}
  \frac{[1+\lambda(\mathbf{m}_{1}\cdot\mathbf{m}_{2})^{2}]g^{*}}{(g_{\rm r}+g^{*})[1-\lambda^{2}(\mathbf{m}_{1}\cdot\mathbf{m}_{2})^{2}]}
  \tanh^{2}
  \left(
    \frac{d_{2}}{2\ell}
  \right)
  m_{2x}
  m_{2y}.
  \label{eq:resistivity_new}
\end{split}
\end{equation}

% ===================================================================================================================================================================================== %

\begin{figure*}[p]
\centerline{\includegraphics[width=1.0\columnwidth]{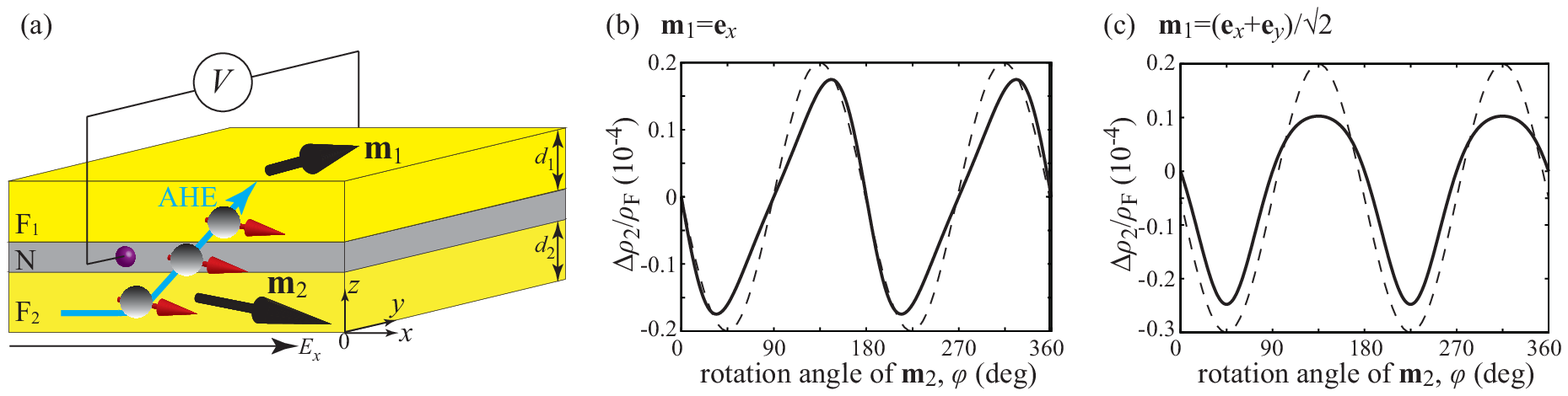}}%\vspace{-3.0ex}
\caption{
        (a) Schematic view of the system under consideration. 
            An external electric field $E_{x}$ is applied along the $x$ direction. 
            The anomalous Hall effect (AHE) in the F${}_{2}$ layer injects spin current into the F${}_{1}$ layer. 
            Transverse magnetoresistance is derived from the voltage generated along the $y$ direction. 
        (b), (c) Dependences of Eq. (\ref{eq:resistivity_new}) on the rotation angle $\varphi$ of the magnetization $\mathbf{m}_{2}=(\cos\varphi,\sin\varphi,0)$ in the F${}_{2}$ layer
            for (b) $\mathbf{m}_{1}=\mathbf{e}_{x}$ and (c) $\mathbf{m}_{1}=(\mathbf{e}_{x}+\mathbf{e}_{y})/\sqrt{2}$. 
            The function $-\sin\varphi\cos\varphi$ is also shown by the dotted line. 
         \vspace{-3ex}}
\label{fig:fig1}
\end{figure*}

The resistivities $\Delta\rho_{1}$ and $\Delta\rho_{2}$ are proportional to the square of the spin anomalous Hall angle defined by Eq. (\ref{eq:spin_Hall_angle}). 
The term $\Delta\rho_{1}m_{2x}m_{2y}$ in Eq. (\ref{eq:total_resistivity}) has the same angular dependence as the planar Hall effect. 
On the other hand, the term $\Delta\rho_{2}(\mathbf{m}_{1},\mathbf{m}_{2})m_{2x}m_{2y}$ depends not only on $m_{2x}m_{2y}$ 
but also on $(\mathbf{m}_{1}\cdot\mathbf{m}_{2})^{2}$. 
Figures \ref{fig:fig1}(b) and \ref{fig:fig1}(c) show examples of $\Delta\rho_{2}/\rho_{\rm F}$ 
as a function of the rotation angle $\varphi$ of the magnetization $\mathbf{m}_{2}=(\cos\varphi,\sin\varphi,0)$ in the F${}_{2}$ layer, 
where the magnetization in the other (F${}_{1}$) layer points to the direction (b) $\mathbf{m}_{1}=\mathbf{e}_{x}$ and (c) $\mathbf{m}_{1}=(\mathbf{e}_{x}+\mathbf{e}_{y})/\sqrt{2}$. 
The angular dependence of the planar Hall effect ($\propto m_{2x}m_{2y}=\sin\varphi\cos\varphi$) is also shown by the dotted line as for guide. 
We referred the values of the parameters from typical experiments and calculations; 
$\ell=1.0$ nm, $\rho_{\rm F}=300$ $\Omega$nm, $\beta=0.90$, $r=0.25$ k$\Omega$nm${}^{2}$, $\gamma=0.50$, $\sigma_{\rm AH}/\sigma_{\rm F}=0.015$, $\zeta=1.50$, and $d_{1}=d_{2}=2.0$ nm \cite{kim16,zwierzycki05,bass07}. 
We assume that the parameters are the same between two ferromagnets, except $\sigma_{\rm AH}$ which is zero in the F${}_{1}$ layer, for simplicity. 
It is clearly shown that the resistivity $\Delta\rho_{2}m_{2x}m_{2y}$ has a slightly different angular dependence from the planar Hall effect, 
and depends on the direction of the magnetization $\mathbf{m}_{1}$ in the F${}_{1}$ layer. 
Therefore, the resistivity $\Delta\rho_{2}$ can be distinguished from the planar Hall effect when a material having a large spin anomalous Hall angle is found. 
We also note that the spin Nernst effect \cite{meyer17,Sheng17,SNE3} in nonmagnets provides physical phenomena similar to that originating from the spin Hall effect. 
Therefore, we expect that the magnetoresistance calculated in this paper will be discovered not only by the anomalous Hall effect but also by the anomalous Nernst effect, 
which was recently shown to contribute to the charge-spin conversion in ferromagnets \cite{taniguchi16JPSJ}. 
%The value of $\Delta\rho_{2}/\rho_{\rm F} \sim 10^{-5}$ is one order of magnitude smaller than 
%the spin Hall magnetoresistance in the nonmagnet calculated from typical parameters \cite{chen13}. 
%This is because the spin anomalous Hall angle, $|(\beta-\zeta)\sigma_{\rm AH}/\sigma_{\rm F}|\simeq 0.009$, is much smaller than 
%the spin Hall angle ($\theta=0.05$) used in Ref. \cite{chen13}. 
%As mentioned in the introduction, recently, the experimental efforts have been made to observe the charge-spin conversion by the anomalous Hall effect. 
%A large $\Delta\rho_{2}$ is expected when a material having a large spin anomalous Hall angle will be found through such experiments. 

% ===================================================================================================================================================================================== %

\section{Conclusion}
\label{sec:Conclusion}

In conclusion, 
a theoretical framework on the transverse magnetoresistance in a ferromagnetic/nonmagnetic/ferromagnetic trilayer originating from charge-spin conversion by anomalous Hall effect was developed. 
%An analytical formula of the transverse resistivity was derived. 
%It was clarified that the charge-spin conversion by the anomalous Hall effect gives the magnetoresistance proportional to the square of the spin anomalous Hall angle. 
It was shown that the angular dependence of the magnetoresistance is basically identical that of planar Hall effect, 
but also has an additional term which depends on the relative angle of the magnetizations in two ferromagnets. 

% ===================================================================================================================================================================================== %

\section*{Acknowledgement}

The author is thankful to Shinji Yuasa, Hitoshi Kubota, Akio Fukushima, Kay Yakushiji, Takehiko Yorozu, Yoichi Shiota, Sumito Tsunegi, 
Atsushi Sugihara, Takahide Kubota, Satoshi Iba, Aurelie Spiesser, Hiroki Maehara, and Ai Emura for their support and encouragement. 
This work is supported by JSPS KAKENHI Grant-in-Aid for Young Scientists (B) 16K17486. 

% ===================================================================================================================================================================================== %

% ===================================================================================================================================================================================== %

% ===================================================================================================================================================================================== %

% ===================================================================================================================================================================================== %

% ===================================================================================================================================================================================== %

%\bibliography{biblist}% Produces the bibliography via BibTeX.

% ===================================================================================================================================================================================== %

\end{document}